\documentclass[preprint,amsmath,amssymb,aps,showpacs,showkeys,prd]{revtex4-2}

\usepackage{graphicx}
\usepackage{dcolumn}
\usepackage{bm}
\usepackage{amsmath}
\usepackage{amssymb}
\usepackage{amsfonts}
\usepackage{slashed} 
\usepackage{comment}
\usepackage{color}
\usepackage{ulem}
\usepackage[colorinlistoftodos]{todonotes}

\numberwithin{equation}{section}

\begin{document}

\preprint{BARI-TH/758-24}

\title{QCD with (2+1) flavors at the physical point in external chromomagnetic fields}

\author{Paolo Cea }
\email{paolo.cea@ba.infn.it}
\author{Leonardo Cosmai}
 \email{leonardo.cosmai@ba.infn.it}
\affiliation{ INFN - Sezione di Bari - Italy }

\begin{abstract}
We investigate  full QCD with (2+1)-flavour of HISQ fermions at the physical point in   the presence of  uniform 
Abelian chromomagnetic background  fields.  Our focus is on the renormalized light and strange chiral condensate around the 
pseudo-critical temperature. We find that in the confined region the gauge system is subjected to the chromomagnetic catalysis 
that turns into the inverse catalysis in the high-temperature regime.  We further observe that the chiral condensates are subjected to 
the so-called thermal hysteresis. Our estimate of the deconfinement temperature indicates that  the critical temperature begins to decrease
 in the small field region, soon after  it seems to saturate  and finally increase with the strength of the chromomagnetic field.
\end{abstract}




\maketitle

\newpage

\section{Introduction}
\label{S1}
Quantum Chromodynamics (QCD) is the theory of quarks and gluons whose interaction is described by a local SU(3)
gauge symmetry~\cite{gross:2022hyw}. Recently, the properties of strong interactions in presence of external fields
have attracted great interests (for a review see~\cite{Endrodi:2024cqn}).  In particular, large magnetic or chromomagnetic background fields are of phenomenological
and theoretical importance since they are able to affect the transition of strongly interacting matter from the
deconfined to the confined phase and, thereby, can provide useful informations on the structure of the QCD vacuum. \\
QCD with external background magnetic fields has been studied extensively using non-perturbative lattice
simulations~\cite{Kharzeev:2013,Andersen:2014xxa}. Indeed, by means of improved staggered quarks with physical masses
the QCD phase diagram has been mapped for magnetic fields up to field strength $\sqrt{eB}$ = 3.0 GeV
(see Refs.~\cite{Bali:2011qj,Bali:2012zg,Endrodi:2015oba,DElia:2021yvk} and references therein).
These studies suggest that the magnetic field enhances the light quark condensate in the confined phase, leading to the phenomenon of the so-called magnetic catalysis. On the other hand, in the transition region from the confined phase to the deconfined one the
contribution of the magnetic fields begins to decrease leading to the inverse magnetic catalysis in the
high-temperature region. As a result of the non-monotonous dependence of the light quark condensate,
the critical deconfinement temperature (assumed to be coincident with the chiral pseudocritical temperature) 
turns out to be reduced by the applied magnetic field suggesting that it tends toward zero for
extremely magnetic field strengths $\sqrt{eB} \ge 2$ GeV~\cite{DElia:2021yvk}. \\
This paper presents an exploratory study of the quark chiral condensate and deconfinement critical temperature in full QCD with 2+1 flavors at the physical point, under the influence of uniform Abelian chromomagnetic fields.
There is an important difference between the case of magnetic and chromomagnetic fields. In fact, a magnetic field
belongs to the electromagnetic U(1) gauge symmetry group such that it can be coupled only to quarks and, therefore,
does not interfere with updating of the SU(3) degrees of freedom during Monte Carlo simulations. On the other hand,
chromomagnetic background fields are built of with the same degrees of freedom which are dynamically updated
during numerical simulations. 

Fortunately, a method for implementing static Abelian chromomagnetic fields on the lattice was established long ago. This method utilizes the so-called Schr\"odinger functional for both Abelian and non-Abelian gauge theories~\cite{Cea:1997ff,Cea:1997ku,Cea:1999gn}.\\
The present paper is organised as follows.  In Sect.~\ref{S2}  we briefly discuss the Schr\"odinger functional
and the Abelian background chromomagnetic field on the lattice.
In Sect.~\ref{S3} we illustrate our lattice setup and give some details on our numerical simulations.
Section~\ref{S4} is devoted to present our numerical results for the dimensionless renormalized 
chiral condensate for both light and strange quarks. Here, we also report our findings on the determination of the pseudocritical temperatures. In Sect.~\ref{S5} we present the behaviour of the pseudocritical temperatures as a
function of the chromomagnetic field with strengths ranging in the interval 0.49 GeV $\lesssim \sqrt{gH} \lesssim$
2.63 GeV. Finally, in Sect.~\ref{S6} we summarise the main results and sketch our conclusions. 
\section{Abelian chromomagnetic fields on the lattice}
\label{S2}
In Refs.~\cite{Cea:1997ff,Cea:1997ku,Cea:1999gn} it was introduced on the lattice
the  zero-temperature  gauge invariant effective action $\Gamma[\vec{A}^{\text{ext}}]$ for an external
background field $\vec{A}^{\text{ext}}$:
\begin{equation}
\label{2.1} 
\Gamma[\vec{A}^{\text{ext}}] = -\frac{1}{L_t} \ln
\left\{
\frac{{\mathcal{Z}}[\vec{A}^{\text{ext}}]}{{\mathcal{Z}}[0]}
\right\} \; .
\end{equation}
where $L_t$ is the lattice size in time direction and $\vec{A}^{\text{ext}}(\vec{x})$  is the continuum gauge potential for the 
external static background field.  The corresponding lattice links are given by:
\begin{equation}
\label{2.2}
U^{\text{ext}}_k(\vec{x})= P \exp \int_0^1 dt \,\, iag A^{\text{ext}}_k(\vec{x}+ta\hat{k})
\end{equation}
where $a$ is lattice spacing, $g$ the bare gauge coupling and  $P$ the path ordering operator. In Eq.~(\ref{2.1})
${\mathcal{Z}}[\vec{A}^{\text{ext}}]$ is the following lattice partition functional (Schr\"odinger functional):
\begin{equation}
\label{2.3} 
{\mathcal{Z}}[\vec{A}^{\text{ext}}] = 
\int_{U_k(\vec{x},x_t=0)=U_k^{\text{ext}}(\vec{x})}
{\mathcal{D}}U \; e^{-S_G} \,, 
\end{equation}
where $S_G$ is the gauge action.  The functional integrations are performed over the lattice links where 
the spatial links belonging to a given time slice  $x_t=0$ are constrained to:
\begin{equation}
\label{2.4}
U_k(\vec{x},x_t=0) = U^{\text{ext}}_k(\vec{x})
\,,\,\,\,\,\, (k=1,2,3) \,\,,
\end{equation}
$U^{\text{ext}}_k(\vec{x})$ being the lattice version of the external  continuum gauge field 
$\vec{A}^{\text{ext}}(x)=\vec{A}^{\text{ext}}_a(x) \lambda_a/2$,  $\lambda_a/2$ being the Gell-Mann matrices. 
Note that the temporal links are not constrained. Since we are interested in the case of a static constant background  fields{\color{red},} we must also impose
that spatial links exiting from sites belonging to the spatial boundaries  are fixed
according to Eq.~(\ref{2.4}).  In the continuum this last condition amounts to the usual requirement that fluctuations over the
background field vanish at spatial infinity.
The lattice effective action $\Gamma[\vec{A}^{\text{ext}}]$ corresponds to the vacuum
energy, $E_0[\vec{A}^{\text{ext}}]$,
in presence of the background field with respect to the vacuum energy, $E_0[0]$, with  $\vec{A}^{\text{ext}}=0$:
\begin{equation}
\label{2.5}
\Gamma[\vec{A}^{\text{ext}}] \quad \longrightarrow \quad E_0[\vec{A}^{\text{ext}}]-E_0[0] \,.
\end{equation}
The previous  relation  holds when   $L_t \to \infty$. On a finite lattice this amounts to have $L_t$ sufficiently large 
to single out the ground state contribution to the energy. \\
The extension to finite temperatures is straightforward. Indeed, if we now consider the gauge theory at finite temperature $T=1/(a L_t)$ 
in presence of an external background field, the relevant quantity turns out to be the free energy functional:
\begin{equation}
\label{2.6}
{\mathcal{F}}[\vec{A}^{\text{ext}}] = -\frac{1}{L_t} \ln
\left\{
\frac{{\mathcal{Z_T}}[\vec{A}^{\text{ext}}]}{{\mathcal{Z_T}}[0]}
\right\} \; .
\end{equation}
${\mathcal{Z_T}}[\vec{A}^{\text{ext}}]$ is the thermal partition functional~\cite{Gross:1981br}
in presence of the background field $\vec{A}^{\text{ext}}$, given by:
\begin{equation}
\label{2.7} 
\mathcal{Z}_T \left[ \vec{A}^{\text{ext}} \right]
= \int_{U_k(\vec{x},L_t)=U_k(\vec{x},0)=U^{\text{ext}}_k(\vec{x})}
\mathcal{D}U \, e^{-S_G}   \,,
\end{equation}
If the physical temperature is sent to zero, the thermal functional, Eq.~(\ref{2.7}), reduces to the zero-temperature Schr\"odinger functional, Eq.~(\ref{2.3}).
Moreover, the free energy functional, Eq.~(\ref{2.6}), corresponds  to the free energy in presence
of the external background field evaluated with respect  to the free energy without external field. \\
When including dynamical fermions, the thermal partition functional
in presence of a static external background gauge field, Eq.~(\ref{2.7}), becomes:
\begin{eqnarray}
\label{2.8}
\mathcal{Z}_T \left[ \vec{A}^{\text{ext}} \right]  &=& 
\int_{U_k(L_t,\vec{x})=U_k(0,\vec{x})=U^{\text{ext}}_k(\vec{x})}
\mathcal{D}U \,  {\mathcal{D}} \psi  \, {\mathcal{D}} \bar{\psi} e^{-(S_G+S_F)} 
\nonumber \\ 
&=&  \int_{U_k(L_t,\vec{x})=U_k(0,\vec{x})=U^{\text{ext}}_k(\vec{x})}
\mathcal{D}U e^{-S_G} \, \det M \,,
\end{eqnarray}
where $S_F$ and $M$ are  the fermionic action and  the fermionic matrix respectively.
Notice that the fermionic fields are not constrained and the integration constraint is only relative to the gauge fields.
As a consequence, this leads, as in the usual QCD partition function, to the appearance of  the gauge invariant fermionic determinant after
 integration on the  fermionic fields.  As usual we impose on fermionic fields periodic boundary conditions in the spatial directions and
antiperiodic boundary conditions in the temporal direction. \\
We are interested in  simulations of lattice QCD with 2+1 flavours of rooted staggered 
quarks.  This means that the thermal  partition functional  of the discretised theory reads:
\begin{equation}
\label{2.9}
\mathcal{Z}_T \left[ \vec{A}^{\text{ext}} \right]  =  
 \int_{U_k(L_t,\vec{x})=U_k(0,\vec{x})=U^{\text{ext}}_k(\vec{x})}
\mathcal{D}U e^{-S_G} \, 
\prod_{f=u,d,s} {\rm det}(D_f[U,m_f])^{1/4} \; .
\end{equation}
In Eq.~(\ref{2.9})  $S_{G}$ will be the Symanzik tree-level improved gauge  action and $D_f[U,m_f]$  the staggered Dirac operator.
As is well known,  the fourth root of the staggered fermion determinant is necessary to represent the contribution of a single fermion
flavor to the QCD partition function. \\
We consider a static constant Abelian chromomagnetic field that in the continuum  is given by:
\begin{equation}
\label{2.10}
\vec{A}^{\mathrm{ext}}_a(\vec{x}) =
\vec{A}^{\mathrm{ext}}(\vec{x}) \,  \delta_{a,3} \,, \quad
A^{\mathrm{ext}}_k(\vec{x}) =  \delta_{k,2} \,  x_1 H \,.
\end{equation}
The corresponding  SU(3) lattice links are:
\begin{equation}
\label{2.11}
\begin{split}
& U^{\mathrm{ext}}_1(\vec{x}) =
U^{\mathrm{ext}}_3(\vec{x}) = {\mathbf{1}} \,,
\\
& U^{\mathrm{ext}}_2(\vec{x}) =
\begin{bmatrix}
\exp(i \frac {gH \, x_1} {2})  & 0 & 0 \\ 0 &  \exp(- i \frac {gH
\, x_1} {2}) & 0
\\ 0 & 0 & 1
\end{bmatrix}
\end{split}
\end{equation}
To be consistent with the hypertorus geometry one should   impose that the
chromomagnetic field is quantized according to:
\begin{equation}
\label{2.12} 
\frac{ a^2 \, gH}{2} = \frac{2 \pi}{L_s} \; 
n_{ext} \,, \qquad  n_{ext}\,\,\,{\text{integer}} \; 
\end{equation}
where the lattice spatial volume is $L_s^3$. However, it should be kept in mind that we are dealing with a periodic
lattice with fixed boundary conditions, so that is is not strictly necessary to impose the
quantisation Eq.~(\ref{2.12}). In the following we shall parametrise  the strength of the chromomagnetic field according to
Eq.~(\ref{2.12}), but we shall keep in mind that  the "integer"  $n_{ext}$ can be an arbitrary real number.
\section{The lattice setup}
\label{S3}
We performed simulations of lattice QCD with 2+1 flavours of highly improved rooted staggered quarks (HISQ). We have made use of
the HISQ/tree action as implemented in the publicly available MILC code~\cite{MILC}. We have modified the MILC code to introduce the cold wall at $x_t=0$ and the fixed spatial boundary conditions.
All simulations were performed using the rational hybrid Monte Carlo (RHMC) algorithm. The length of each RHMC trajectory
has been set to 1.0 in the molecular time units. The functional integrations are done over the SU(3) lattice links
taking fixed according to Eq.~(\ref{2.4}) the links at the spatial boundaries. Accordingly, these frozen links are not
evolved during the molecular dynamics so that the corresponding conjugate momenta are set to zero. \\
For each value of the gauge coupling $\beta$ and chromomagnetic field $gH$ we have simulated the theory on lattices $L_s^3 \times L_t$
discarding a few hundred trajectory to allow thermalization and collected 500 - 1000 trajectories for measurements.
\begin{figure}[t]
\includegraphics[width=0.95\textwidth,clip]{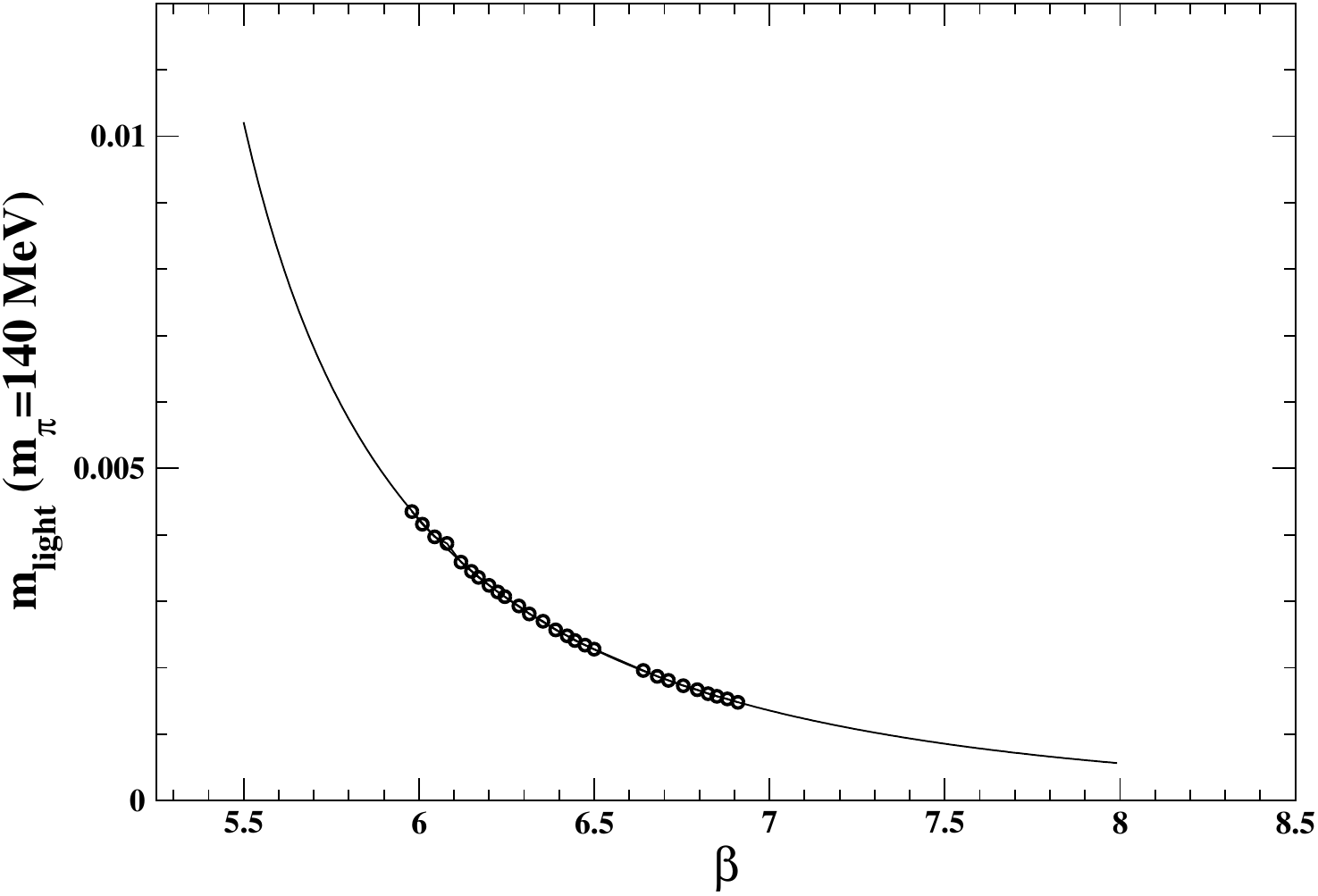}
\caption{\label{Fig1} 
The lattice light  quark mass along the line of constant physics. The lattice data have been taken from Table IV in 
Ref.~\cite{Bazavov:2017dus}. The solid line is the best fit to Eq.~(\ref{3.1}).}
\end{figure}
\begin{table}[t]
\caption{
\label{Table1}
Ensemble parameters for calculations with light to strange quark mass $\frac{m_{\ell}}{m_s}$  $ = $ $\frac{1}{27}$
on lattices of size $L_s^3$  $\times$ $L_t$.
}
\begin{ruledtabular}
\begin{tabular}{ccccccc}
 $\beta$   & $L_s$ & $n_{ext}$ & $a(\beta) \; (fm)$ &
$\sqrt{gH}$ (GeV)  \\
\hline
6.000 & 36 & 1.000 & 0.2379 & 0.4901   
  \\
5.900  & 36 & 1.239 & 0.2648 & 0.4901   
 \\
6.300  & 36 & 0.528 & 0.1728 & 0.4901   
 \\
6.000 & 36 & 2.000 & 0.2379 & 0.6930 
  \\
6.280 & 36 & 1.103 & 0.1765 & 0.6930  
  \\
6.300 & 36 & 1.106 & 0.1728 & 0.6930 
 \\
6.280 & 36 & 3.000 & 0.1765 & 1.1439
 \\
6.100 & 36 & 4.398 & 0.2137 & 1.1439 
 \\
6.300 & 36 & 2.876 & 0.1728 & 1.1439
 \\
6.300  & 36 & 4.000 & 0.1728  &  1.3489  
 \\
6.500 & 36 & 4.000 & 0.1404 & 1.6605 
  \\
 6.640 & 36 & 3.012 & 0.1219 & 1.6605   
  \\
6.850  & 48 & 3.000 & 0.0991 & 1.7644
 \\
6.850 & 36 & 3.000 & 0.0991 & 2.0373  
 \\
6.850 & 36 & 5.000 & 0.0991 & 2.6301 
 \\
\end{tabular}
\end{ruledtabular}
\end{table}
%
The measurement of the chiral condensate has been done using the noisy estimator approach with 12 random vectors. But we have checked that results are consistent if we use 100 random vectors.
The calculations are performed using a strange quark mass $m_s$ tuned to the physical value and light to strange quark mass ratio
$\frac{m_{\ell}}{m_s} = \frac{1}{27}$ with $m_u = m_d = m_{\ell}$ corresponding to continuum limit pion mass 
$m_{\pi} \simeq 140$ MeV~\cite{Bazavov:2010ru,Bazavov:2011nk,Bazavov:2017dus,HotQCD:2018pds}.
The parameters are fixed using the line of constant physics determined from the $f_K$ scale allowing us to evaluate the lattice
spacing $a(\beta)$ at a given value of the gauge coupling, the set of quark masses $(m_{\ell}, m_s)$ and the temperature
$T = [ L_t a(\beta)]^{-1}$. The line of constant physics is defined by $\frac{m_{\ell}}{m_s} = \frac{1}{27}$ and the light quark mass in unit
of the lattice spacing tuned to:
\begin{equation}
\label{3.1}
m_l(\beta) =   \frac{a_1+a_2\beta+a_3\beta^2+a_4\beta^3}{1+a_5\beta+a_6\beta^2+a_7\beta^3}\; .
\end{equation}
The parameters in Eq.~(\ref{3.1}) are fixed by fitting this last equation to the data reported in Table IV of Ref.~\cite{Bazavov:2017dus}.
From the best-fit we obtained (see Fig.~\ref{Fig1}):
\begin{eqnarray}
\label{3.2}
& a_1 =16363.060237188531; \;  a_2=-1004.7534566983386; \;   
\\ \nonumber
& a_3 =-228.15737907741254; \; a_4=18.412492683609351; 
\\ \nonumber
& a_5 =-1142594.5815322001; \; a_6=230401.69003500676; 
\\ \nonumber 
& a_7 =1.2439840951701624 \;.
\end{eqnarray}
The lattice spacing in physical units is given by~\cite{Bazavov:2011nk,MILC:2010hzw}:
\begin{equation}
\label{abeta}
a(\beta) \;=\; \frac{r_1}{r_1 f_K} \frac{c_0^K f(\beta)+c_2^K (10/\beta) f^3(\beta)}{
1+d_2^K (10/\beta) f^2(\beta)} 
\end{equation}
where:
\begin{equation}
\label{r1fkappa}
\begin{split}
r_1 \;  &= \; 0.3106 \; \mathrm{fm}
\\
 r_1 f_K \;  &= \; 
\frac{0.3106 \; \mathrm{fm} \; \cdot \; 156.1/\sqrt{2} \; \mathrm{MeV}}{
197.3 \; \mathrm{MeV \; fm}} \; \;
\end{split}
\end{equation}
and
 \begin{equation}
\label{coefficients}
c_0^K = 7.66 \; \;  , \; \;  c_2^K \; = \; 32911 \; \; , \; \;  d_2^K  \;  =  \;  2388 \; \; .
\end{equation}
In Eq.~(\ref{abeta})  $f(\beta)$ is the two-loop beta  function:
\begin{equation}
\label{3.6}
f(\beta)=[b_0 (10/\beta)]^{-b_1/(2 b_0^2)} \exp[-\beta/(20 b_0)] \; ,
\end{equation}
$b_0$ and $b_1$ being its universal coefficients. \\
The simulation parameters are summarised in Table~\ref{Table1}.
\section{The renormalized chiral condensate}
\label{S4}
As discussed in Sect.~\ref{S1} one of the most remarkable effect of a magnetic background field on the QCD vacuum
manifests itself as the magnetic catalysis in the confined phase, namely the enhancement of the quark chiral condensate.
On the other hand,  in the deconfined region there is ample evidence of the inverse magnetic catalysis leading to a decrease in the critical temperature with increasing magnetic field strength. 
The mechanisms responsible for the magnetic catalysis are believed to originate from the dimensional reduction
of dynamical quarks in presence of strong magnetic fields. Indeed, the enhancement of the chiral condensate at low temperatures 
should arise naturally from the fact that quarks are frozen into the lowest Landau levels 
(a good account can be found in Refs.~\cite{Endrodi:2024cqn,Miransky:2015ava,Hattori:2023egw} and references therein).
However, it should be mentioned that, up to now, theoretical attempts to uncover the magnetic catalysis mechanism are in 
reasonable agreement with the non-perturbative lattice data only for not too strong magnetic fields. \\
In lattice simulations the quark condensate is defined as the derivative of the logarithm of the thermal partition functional with respect to the 
quark mass:
\begin{equation}
\label{4.1}
< \bar{\psi} \, \psi > _f(T, gH) \; = \; \frac{T}{V} \; \frac{\partial \, \ln \mathcal{Z}_T \left[ \vec{A}^{\text{ext}} \right]}{\partial \, m_f} \; \; \; ,
\; \; \; f \, = u, \, d, \, s  \; \; 
\end{equation}
where  $V = L_s^3$ is the spatial volume. However, the logarithm of the partition functional contains additive divergences~\cite{Leutwyler:1992yt}
that need to be renormalised. Following Refs.~\cite{Borsanyi:2010bp,Bali:2011qj} we focus on the dimensionless renormalized
chiral condensate:
\begin{equation}
\label{4.2}
\Sigma_f(T,gH) \; = \;  \frac{m_f}{m_{\pi}^4} \; \left [ < \bar{\psi} \, \psi > _f(T, gH) \; - \; < \bar{\psi} \, \psi > _f(0, 0) \right ] \; 
\end{equation}
where $m_{\pi}$ is the $T=0$ pion mass and the multiplication by $m_f$ eliminates multiplicative divergences caused by the
derivative with respect to the quark mass. 
Note that the normalisation in Eq.~(\ref{4.2}) can be easily converted into the slightly different one used in Ref.~\cite{Bali:2012zg}.
These normalisations of the chiral condensate allow to take safely the continuum limit~\cite{Aoki:2006we}.
In the present paper we consider the light chiral condensate:
\begin{equation}
\label{4.3}
\Sigma_{\ell}(T,gH) \; = \;  \frac{1}{2} \; \left \{ \Sigma_u(T,gH) \; + \;  \Sigma_d(T,gH) \right \} 
\end{equation}
and the strange chiral condensate
\begin{equation}
\label{4.4}
\Sigma_s(T,gH) \; = \;  \frac{m_s}{m_{\pi}^4} \; \left [ < \bar{\psi} \, \psi > _s(T, gH) \; - \; < \bar{\psi} \, \psi > _s(0, 0) \right ] \; .
\end{equation}
It is well established that for physical quark masses the transition between the hadronic and the quark-gluon phases is an analytic
crossover~\cite{Aoki:2006we,Bhattacharya:2014ara}. In an analytic transition, due to the absence of singular behaviour,  a unique critical temperature cannot be defined, however,  a pseudocritical temperature  can be determined using the inflexion point or peak position
of  certain thermodynamic observables. Usually, the order parameters employed for chiral and deconfinement phase transitions are the
renormalised chiral condensate and Polyakov loop. The chiral and deconfinement pseudocritical temperatures can be estimated
as the inflection points of the renormalised  chiral condensate of light quarks and the renormalised Polyakov loop respectively.
It turned out that these critical temperatures are consistent even in presence of strong external magnetic fields. Therefore, in the following
we shall assume that the pseudocritical deconfinement temperature always coincides with the light chiral critical temperature for
all the explored values of the chromomagnetic field strength. \\
%
\begin{table}[hbt]
\caption{
\label{Table2}
Values of the parameters from fitting the light chiral condensate to Eq.~(\ref{4.6}). The last column contains the
fitting temperature interval.}
\begin{ruledtabular}
\begin{tabular}{ccccccc}
 $\sqrt{gH}$  (GeV)  & $\Sigma_0^{\ell}$ & $\Delta \Sigma_{\ell}$ & $T_c^{\ell}$ &
$\Delta T_{\ell}$ (MeV) & $T_1$ , $T_2$  (MeV) \\
\hline
0.4901 &- 0.14013$\pm$0.01090 & -0.15575$\pm$0.01833  & 111.2$\pm$4.8 & 59.0$\pm$7.5 &  50 , 140
  \\
0.6930 & -0.12080$\pm$0.01709 & -0.14895$\pm$0.02508 & 125.2$\pm$7.2 & 56.6$\pm$9.5 &  50 , 140 
 \\
1.1439 & -0.05181$\pm$0.00347 & -0.15348$\pm$0.00949 & 129.0$\pm$2.0 & 51.3$\pm$6.0  &  50 , 190
 \\
1.3489 & -0.01110$\pm$0.01857 & -0.16550$\pm$0.05649 & 128.7$\pm$7.7 & 63.5$\pm$29.5  & 90 , 190
  \\
 1.6605 & -0.02495$\pm$0.00344 & -0.13604$\pm$0.00638 & 146.5$\pm$2.6 & 56.1$\pm$5.9  & 78 , 235\footnotemark[1]
  \\
1.6605 & +0.03236$\pm$0.00485 & -0.12163$\pm$0.00646 & 148.9$\pm$2.7 & 32.9$\pm$5.8  & 100 , 250\footnotemark[2]
 \\
1.7644 & +0.01846$\pm$0.00900 & -0.15236$\pm$0.01104 & 133.6$\pm$4.8 & 53.0$\pm$6.2  & 82 , 250
 \\
2.0373 & +0.020861$\pm$0.01887 & -0.09038$\pm$0.02850 & 142.6$\pm$11.2 & 37.0$\pm$18.7  & 100 , 200
 \\
2.6301 & -0.05160$\pm$0.00294 & -0.05987$\pm$0.00428 & 166.0$\pm$3.8 & 36.5$\pm$7.9 & 90 , 250
 \\
\end{tabular}
\end{ruledtabular}
\footnotetext[1]{lower branch}
\footnotetext[2]{upper branch}
\end{table}
\begin{figure}[hbt]
\includegraphics[width=0.95\textwidth,clip]{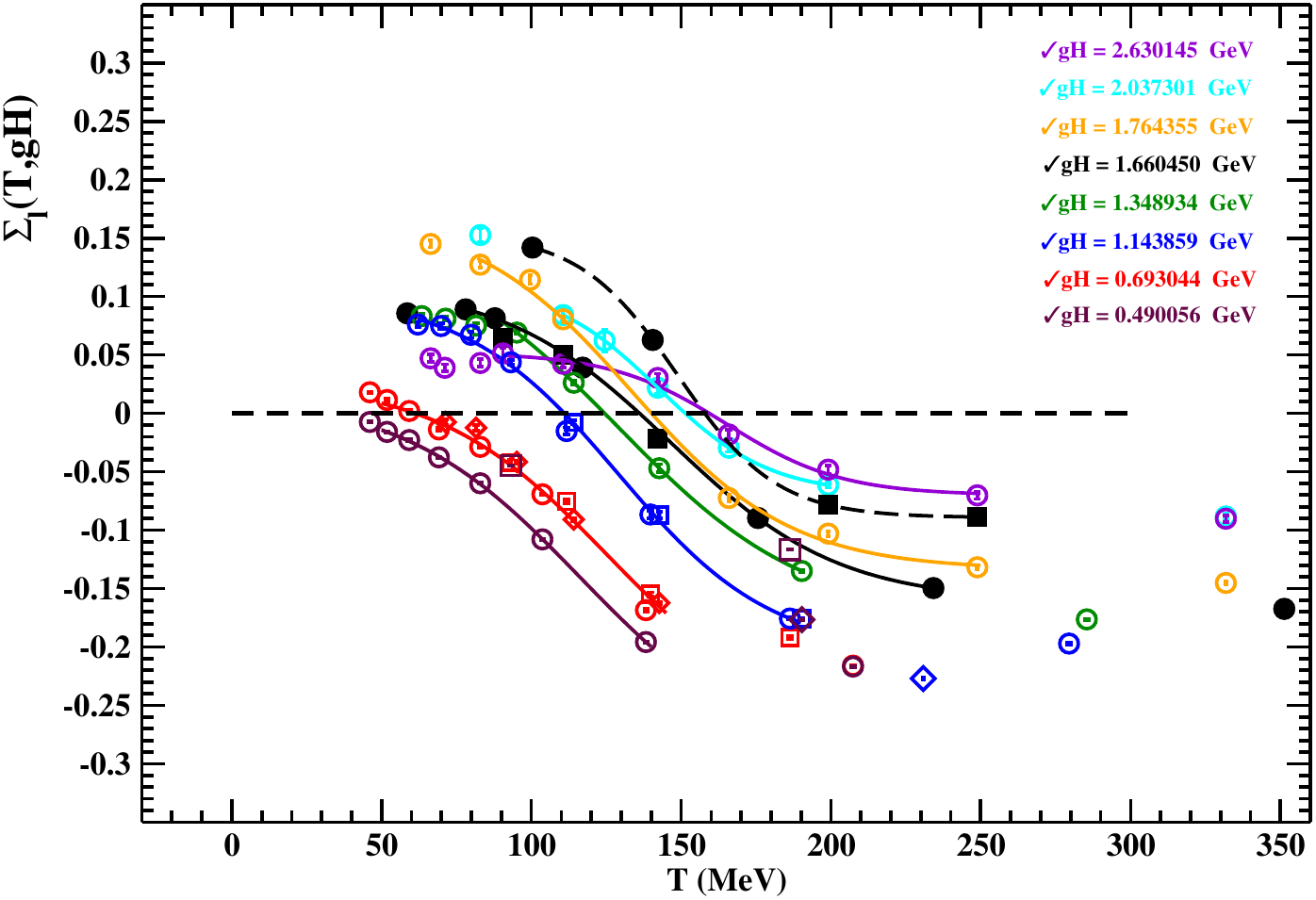}
\caption{\label{Fig2} 
The normalised light chiral condensate versus the temperature for different values of the chromomagnetic field strength. 
The values of the chromomagnetic field strength  are identified by different colors as displayed in the legend.
The maroon circles, squares and 
diamonds correspond to $\beta$ = 6.000, 5.900 and 6.300, the red circles, squares and diamonds to $\beta$ = 6.000, 6.280 and 6.300,
the blu circles, squares and diamonds to $\beta$=6.280, 6.100 and 6.300. 
The full black circles and squares correspond to  $\beta$=6.500  and $\beta$=6.640 respectively.
The continuous lines are the best fits of the lattice data to Eq.~(\ref{4.6}) in the displayed temperature range.}
\end{figure}
%
To determine the critical temperatures we have explored a wide temperature region around the zero-field pseudocritical 
temperature~\cite{HotQCD:2018pds}:
\begin{equation}
\label{4.5}
 T_c \; = \; 156.5 \; \pm \; 1.5 \; \; {\text{MeV}} \;
\end{equation}
for various values of the chromomagnetic field ranging from $\sqrt{gH} \, \simeq \, 0.49$ GeV to $\sqrt{gH} \, \simeq \, 2.63$ GeV.
The pseudocritical temperatures have been determined by fitting the renormalised chiral condensate in the critical region
according to:
\begin{equation}
\label{4.6}
\Sigma_{\ell}(T,gH) \; = \; \Sigma_0^{\ell}(gH) \; + \; \Delta  \Sigma_{\ell}(gH) \; \tanh \left [ \frac{T \; - \; T_c^{\ell}(gH)}{\Delta T_{\ell}(gH)}
\right ] \; \; . 
\end{equation}
The fit to Eq.~(\ref{4.6}) is inspired to the mean field solution of the Ising model where the magnetisation is related to the effective
magnetic field $h_{\text{eff}}$ felt by the spins by:
\begin{equation}
\label{4.7}
M \; \simeq \; \tanh \left ( \frac{h_{\text{eff}}}{T} \right ) \; .
\end{equation}
We have, also, employed the fitting functional form: 
\begin{equation}
\label{4.8}
\Sigma_{\ell}(T,gH) \; = \; \Sigma_0^{\ell}(gH) \; + \; \Delta  \Sigma_{\ell}(gH) \; \arctan \left [ \frac{T \; - \; T_c^{\ell}(gH)}{\Delta T_{\ell}(gH)}
\right ] \; 
\end{equation}
suggested  in Ref.~\cite{Bonati:2018nut}. As a matter of fact, we have checked that both fits gave compatible results for the pseudocritical
temperature $T_c^{\ell}(gH)$. The main advantage in using Eq.~(\ref{4.6}) resides on the fact that one can directly estimate the
zero-temperature renormalised chiral condensate by:
\begin{equation}
\label{4.9}
\Sigma_{\ell}(T=0,gH) \; \simeq  \; \Sigma_0^{\ell}(gH)\; - \;  \Delta  \Sigma_{\ell}(gH) \; \; .
\end{equation}
In Table~\ref{Table2} we report the results obtained by fitting  our lattice data to Eq.~(\ref{4.6}), while in Fig.~\ref{Fig2} we show our measurements
of the renormalised chiral condensate for light quarks versus the temperature and for eight different values of the chromomagnetic field
strength $\sqrt{gH}$. The measurements have been done along the line of constant physics and constant chromomagnetic field strength.
Most our simulations have been carried out keeping the spatial size and $n_{ext}$ fixed and varying the temporal lattice size $L_t$ in
order to change the temperature. In a few cases, in order to check   possible finite size effects or scaling violations, we varied the spatial
lattice size, the gauge  coupling and the parameter $n_{ext}$ such that the chromomagnetic field is kept fixed in physical
units according to Eq.~(\ref{2.12}).  Looking at Fig.~\ref{Fig2} we can safely affirm that  possible scaling violations are under control and,
therefore, they do not affect significantly  our numerical results. \\
A remarkable aspect of our Fig.~\ref{Fig2} is that the renormalised light chiral condensate seems to behave as   in the case of full QCD in external
magnetic fields. Indeed, at low temperatures, in the confinement phase,  the chiral condensate increases with the chromomagnetic field strength
leading to the chromomagnetic catalysis. On the other hand, the chiral condensate begins to decrease in the deconfined phase
giving rise to the inverse chromomagnetic catalysis. There are, however, important differences as expected since an external
chromomagnetic field couples to both quarks and gluons. Firstly, at low temperature the renormalised chiral condensate in external
magnetic field eB increases monotonically such that  $\Sigma(0,eB)$ behaves almost linearly in eB for strong enough 
magnetic field strengths~\cite{DElia:2021tfb}. 
This large field behaviour of the magnetic catalysis is interpreted theoretically in terms of the lowest Landau level
approximation. In fact, the energy of the lowest Landau level is independent on eB, but its degeneracy linearly
increases with eB. On the other hand, in the low-temperature region a clear theoretical understanding
of the dependence of the renormalised chiral condensate on both the temperature T and the magnetic field eB can be
obtained within the so-called Hadron Resonance Gas model~\cite{Endrodi:2013cs}.
 Basically, the success of the Hadron Resonance Gas model resides on the fact that the external magnetic field couples 
 directly to charged hadrons such that one can easily evaluate the thermodynamic potential by means of the partition function
 of a gas of non-interacting free hadrons and resonances. \\
\begin{figure}
\includegraphics[width=0.95\textwidth,clip]{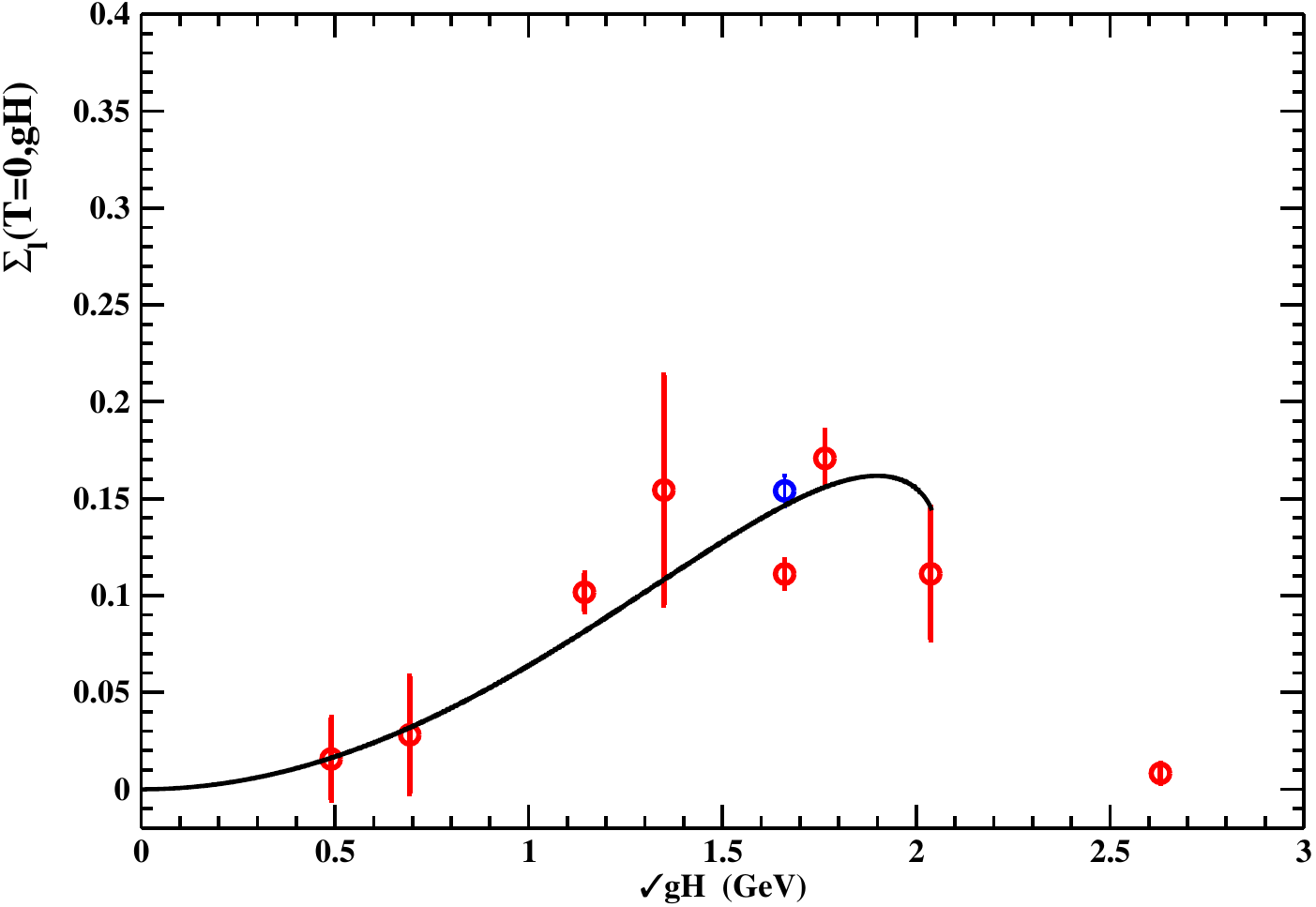}
\caption{\label{Fig3} 
The zero-temperature light chiral condensate given by  Eq.~(\ref{4.9})  versus $\sqrt{gH}$ (red open circles).
The blue open circle refers to the fit of the upper branch at $\sqrt{gH} \simeq 1.66$ GeV in Table~\ref{Table2}.
The continuous line is  Eq.~(\ref{4.11}) multiplied by a magnification factor $A \simeq 15.0$.}
\end{figure}
At variance of the case of external magnetic fields, at low  temperatures Fig.~\ref{Fig2} shows that the light chiral condensate
increases with the strength of the chromomagnetic field $\sqrt{gH}$, reaches a broad maximum and after that it seems to decrease
towards zero for extremely strong chromomagnetic fields. To better appreciate this point, in Fig.~\ref{Fig3} we report the
zero-temperature renormalised light chiral condensate, as estimate from Eq.~(\ref{4.9}), versus $\sqrt{gH}$.
Considering that external chromomagnetic fields cannot couple directly to low-lying vacuum excitations that
are made of colourless hadrons, to understand the peculiar behaviour of the light chiral condensate displayed in Fig.~\ref{Fig3}
we are led to assume that background chromomagnetic fields affect the QCD vacuum such that the low-lying excitation spectrum
gets modified. In the case of the light quark chiral condensate the lowest vacuum excitations are three degenerate pions.
Accordingly, we shall assume that the external chromomagnetic field modifies the pion mass as:
\begin{equation}
\label{4.10}
m^2_{\pi}(gH)  \; \simeq  \; m^2_{\pi} \;  +  \;  \alpha \,  gH \; \; .
\end{equation}
Starting from this last equation and after a suitable additive renormalisation of the vacuum energy density we reach an estimate
of the renormalised light chiral condensate at zero temperature:
\begin{equation}
\label{4.11}
\Sigma_{\ell}(T=0,gH) \; \simeq  \;  \frac{3}{32 \, \pi^2}  \, \left \{ - \alpha \, \frac{gH}{m^2_{\pi}}  \; - \; 
(1 \, + \alpha   \,   \frac{gH}{m^2_{\pi}} ) \ln  (1 \, + \alpha   \,   \frac{gH}{m^2_{\pi}} )     \right \} \; \; .
\end{equation}
By comparing our model to the lattice data, we obtain the following estimate for the parameter $\alpha$: 
\begin{equation}
\label{4.12}
\alpha \; \simeq  \; - \, 0.0045  \; \; .
\end{equation}
Indeed,  after taking into account Eq.~(\ref{4.12}), Eq.~(\ref{4.11}) seems to track quite well the data points but only after a magnification
factor of about 15.0 (see Fig.~\ref{Fig3}). \\
Note that our theoretical estimate extends up to $\sqrt{gH_c} \simeq 2.1$ GeV where $m^2_{\pi}(gH_c) \simeq 0$.   The vanishing pion mass at the critical value $\sqrt{gH_c}$ suggests a loss of pions as quark-antiquark bound states, implying a restoration of chiral symmetry. This last point is consistent with the circumstance that for $\sqrt{gH} >  \sqrt{gH_c}$ the value
of the zero-temperature renormalized chiral condensate is vanishing small.  Clearly, these considerations warrant further investigation through more detailed studies, which fall outside the scope of this exploratory paper. \\
The most dramatic  aspect of our measurements of the renormalised light chiral condensate is due to the presence of thermal hysteresis.
\begin{figure}
\includegraphics[width=0.95\textwidth,clip]{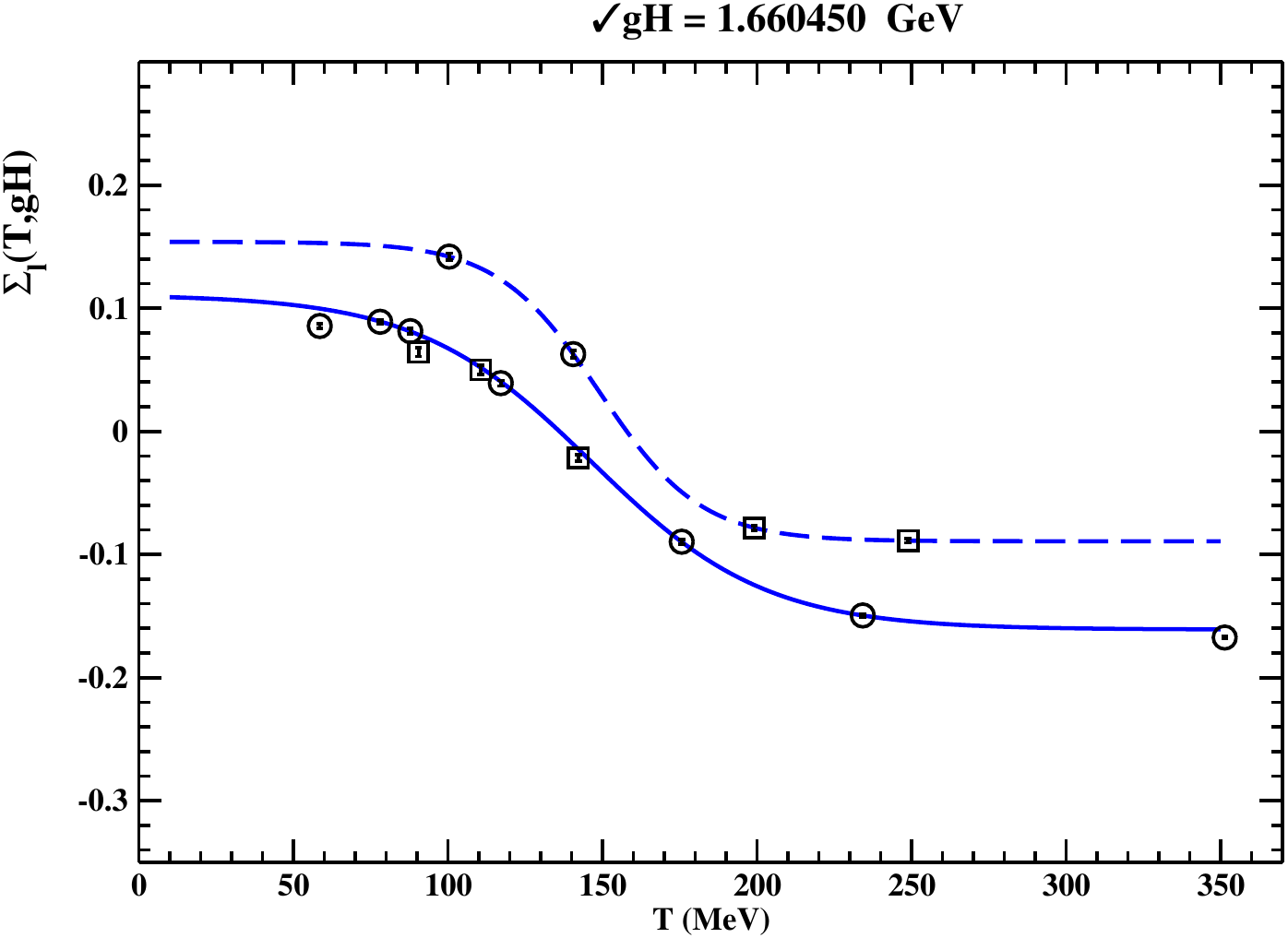}
\caption{\label{Fig4} 
The light chiral condensate at $\sqrt{gH} \simeq	 1.66$ GeV for $\beta =  6.500$ (open circles) and .
 $\beta =  6.640$ (open squares). The full and dashed continuous lines are the fits of the lower and upper branches to  
 Eq.~(\ref{4.6}) respectively.}
\end{figure}
Actually, hysteresis is at the heart of the behaviour of magnetic 
materials~\footnote{A good account can be found in Ref.~\cite{Bertotti:1998} and references therein.}. 
There are several models for the magnetic hysteresis (see, for instance, the recent review Ref.~\cite{Moree:2023} and references therein).
On general grounds,  the thermal hysteresis is due to the existence of metastable free energy minima. As a consequence, the system
is locally brought to an instability by some driving force and then it suddenly moves to a new metastable configuration. In magnetic
systems this kind of behaviour is called a Barkhausen jump~\cite{Bertotti:1998}. Indeed, we found hints of jumps in the renormalised
chiral condensate in the whole explored range of the chromomagnetic field strength. Nevertheless, we found the most clear evidence
of thermal hysteresis at $\sqrt{gH} \simeq 1.66$ GeV. In fact, in Fig.~\ref{Fig4} we report the light chiral condensate at $\sqrt{gH} \simeq 1.66$ GeV
together with the fitting functions extended to the low and high temperature regions. As this figure shows, it seems that hysteresis
extend from zero temperature up to high temperatures. The phenomenon of hysteresis in magnetism is closely tied to the existence of magnetic domains. Indeed,
the free energy has a complicated structure with many extrema and saddle points such that  relaxation towards equilibrium can be an
extremely involved process. In this case, the system is trapped in local energy minima for long times until thermal agitation offers a chance to
overcome one of the energy barriers that separate the system from neighbouring states. In the case of QCD, a straightforward interpretation
of thermal hysteresis would imply the presence of chromomagnetic domains in the QCD quantum vacuum. Interestingly, recent research proposes that at large distances, the confining QCD vacuum behaves like a chromomagnetic condensate characterized by the presence of disordered chromomagnetic domains~\cite{Cea:2023}. However, it must be stressed that hysteresis is a 
challenging phenomena that can occur in rather different situations.
\begin{figure}
\includegraphics[width=0.95\textwidth,clip]{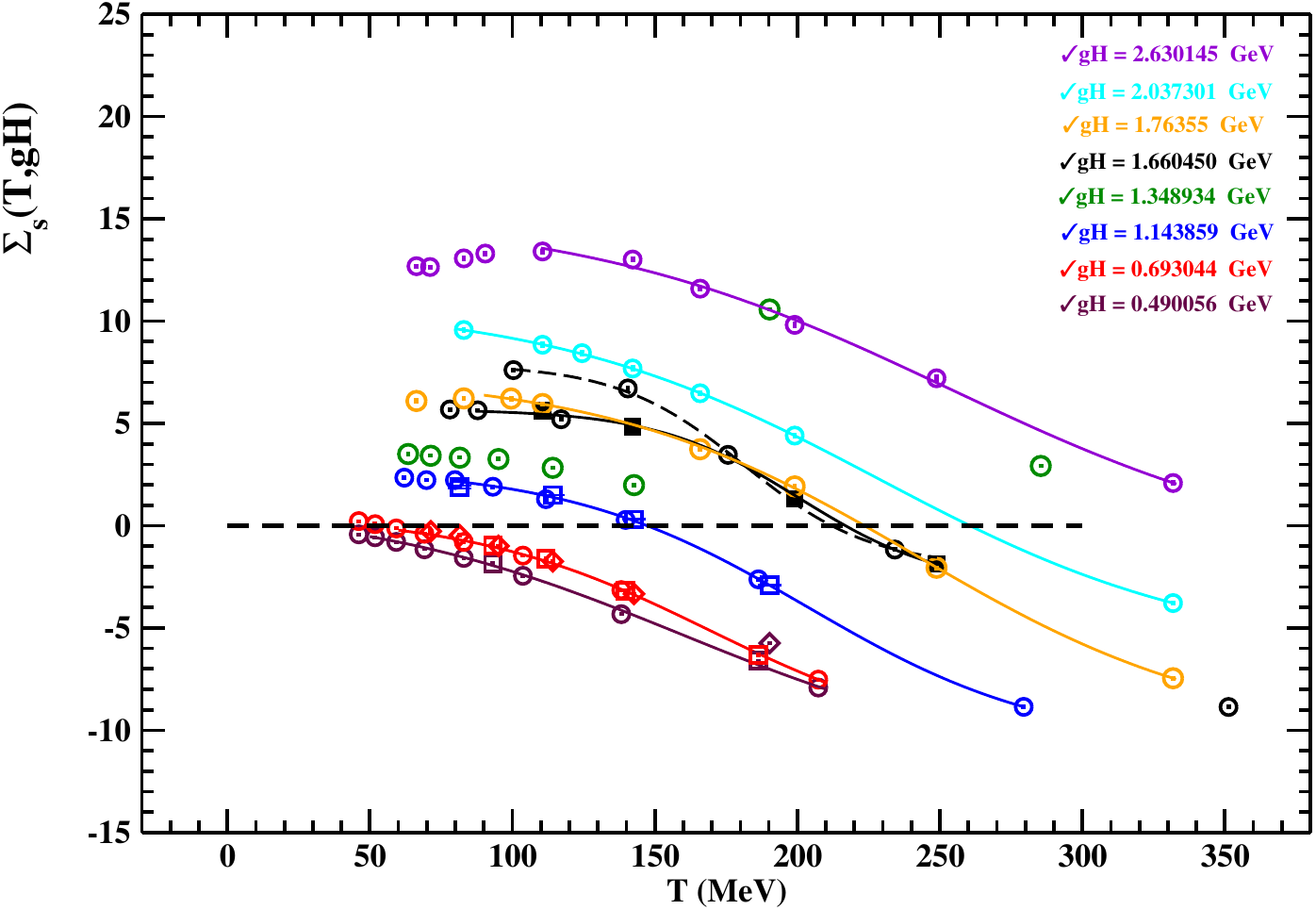}
\caption{\label{Fig5} 
The renormalized strange chiral condensate versus the temperature for different values of the chromomagnetic field strength. 
Symbols as in Fig. \ref{Fig2}. The continuous lines are the  fits  to Eq.~(\ref{4.6}). }
\end{figure}
The observed hysteresis in the thermal behavior of the renormalized chiral condensate could originate from the presence of topological defects in the QCD vacuum~\footnote{For a good overview see, e.g.,  Refs.~\cite{Ripka:2003vv,Greensite:2011zz} and references therein.}.\\
As a further check on the remarkable properties displayed by the renormalised light chiral condensate we have looked also
at the strange chiral condensate as defined by Eq.~(\ref{4.4}). In Fig.~\ref{Fig5} we display the strange chiral condensate as a function of T
for different values of $\sqrt{gH}$.
\begin{table}[t]
\caption{
\label{Table3}
Values of the fitting parameters for the strange chiral condensate. In the last column it is reported the fitting interval.}
\begin{ruledtabular}
\begin{tabular}{ccccccc}
 $\sqrt{gH}$  (GeV)  & $\Sigma_0^{s}$ & $\Delta \Sigma_{s}$ & $T_c^{s}$ &
$\Delta T_{s}$ (MeV) & $T_1$ , $T_2$  (MeV) \\
\hline
0.4901 & -5.30127$\pm$0.11026 & -6.39980$\pm$0.20876 & 159.1$\pm$2.3 & 113.4$\pm$4.1 &  50 , 210
  \\
0.6930& -5.10629$\pm$0.16319 & -5.66771$\pm$0.27069 & 169.1$\pm$2.9 & 83.9$\pm$4.5  &  60 , 210 
 \\
1.1439 & -3.97111$\pm$0.08578 & -6.85689$\pm$0.19543 & 203.5$\pm$1.8 & 85.1$\pm$4.2  &  80 , 280
 \\
 1.6605 & +1.43894$\pm$0.09130 & -4.09855$\pm$0.14908 & 198.9$\pm$1.6 & 45.8$\pm$3.3  & 80 , 250\footnotemark[1]
  \\
1.6605 & +2.94358$\pm$0.07424 & -4.95314$\pm$2.98986 & 181.4$\pm$1.2 & 44.7$\pm$3.0  & 100 , 250\footnotemark[2]
 \\
1.7644 & -1.44039$\pm$0.12761 & -8.86364$\pm$0.33525 & 241.4$\pm$2.1 & 109.3$\pm$6.8  & 90 , 333
 \\
2.0373 & +2.29375$\pm$0.34297 & -8.60607$\pm$0.82560 & 228.5$\pm$8.4 & 117.6$\pm$18.6  & 80 , 330
 \\
2.6301 & +6.71574$\pm$0.53652 & -8.61405$\pm$1.05372 & 253.0$\pm$10.9 & 131.4$\pm$20.6  & 110 , 333
 \\
\end{tabular}
\end{ruledtabular}
\footnotetext[1]{lower branch}
\footnotetext[2]{upper branch}
\end{table}
\begin{figure}
\includegraphics[width=0.95\textwidth,clip]{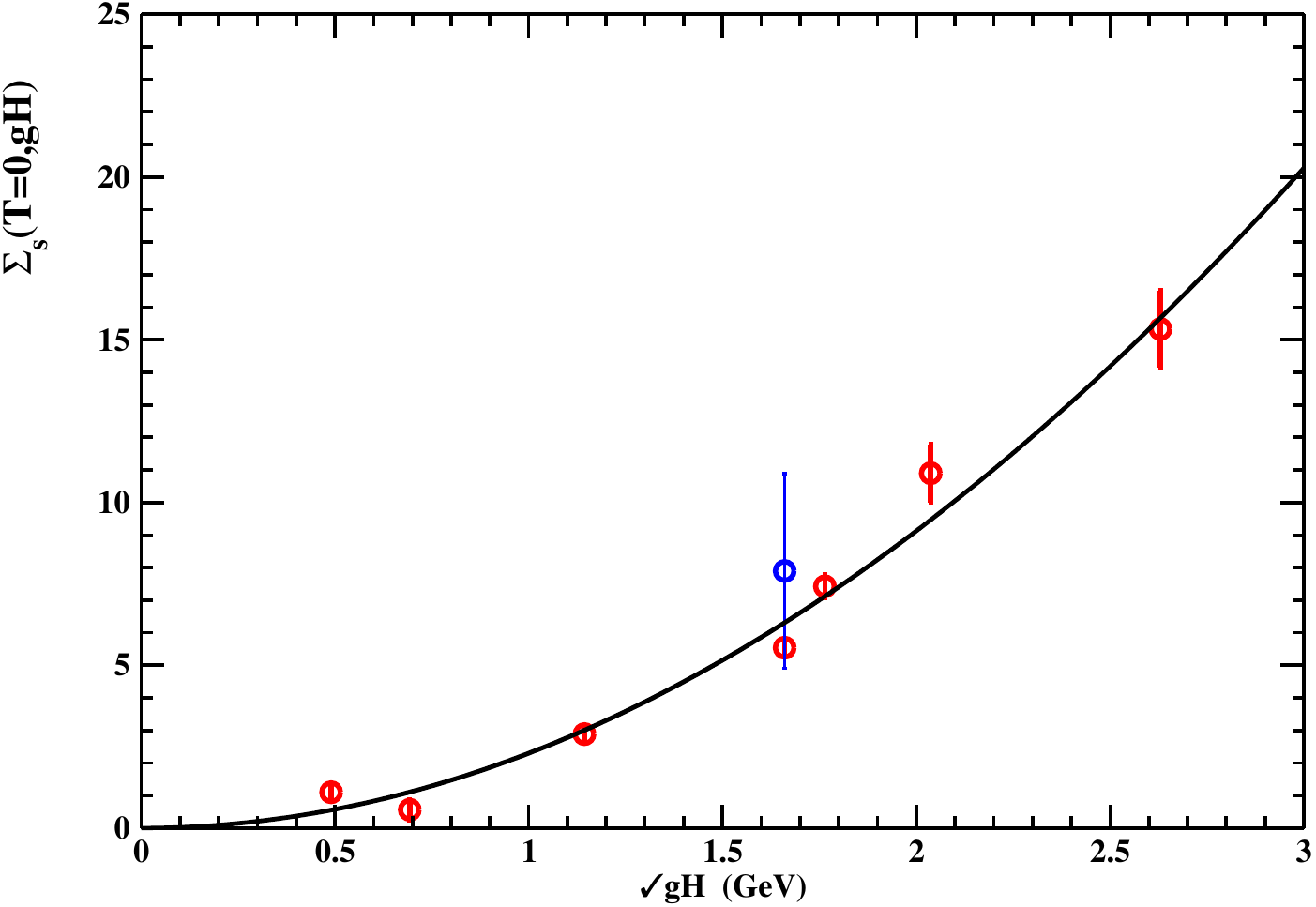}
\caption{\label{Fig6} 
The zero-temperature strange chiral condensate  as a function of $\sqrt{gH}$ (red open circles).
The blu open circle corresponds to   the fit of the upper branch at $\sqrt{gH} \simeq 1.66$ GeV.
The continuous line is  Eq.~(\ref{4.15})  with  the parameter $\alpha$ given by Eq.~(\ref{4.12})
and a magnification factor $A \simeq 66.0$.}
\end{figure}
Even for the strange chiral condensate  the lattice data have been fitted to Eq.~(\ref{4.6}) 
(continuous lines in Fig.~\ref{Fig5}) and the resulting best-fitted parameters are reported in Table~\ref{Table3}.
Note that at    $\sqrt{gH} \simeq 1.35$ GeV the presence of hysteresis did not allow a reliable estimate of the critical temperature.  \\
Figure 5 confirms, as expected,  that the main features of the strange chiral condensate agree with  those of the
light chiral condensate. Again, we see that hysteresis manifests itself for all the explored values of $\sqrt{gH}$.
Even in this case, the most clear evidence of hysteresis happens at $\sqrt{gH} \simeq 1.66$ GeV. However,
in this case the upper and lower branches seem to merge in the high-temperature region.
Also, it is confirmed that the chromomagnetic catalysis at low temperatures turns into the inverse catalysis at high temperatures.
In contrast to the previous case, the zero-temperature strange chiral condensate exhibits a monotonic increase with increasing chromomagnetic field strength  (see Fig.~\ref{Fig6}). If we try to track the lattice data within our previous resonance gas model
we must take into account that for the present case the lowest excitation should be the $s\bar{s}$-pseudoscalar bound state
used by the MILC Collaboration to set the scale of the strange quark mass~\cite{Bazavov:2011nk}:
\begin{equation}
\label{4.13}
m_{s\bar{s}}  \; =  \; \sqrt{2 \, m^2_K \; - \; m^2_{\pi}}  \;  \simeq  \;  686 \; \text{MeV} \;  \; .
\end{equation}
Assuming
\begin{equation}
\label{4.14}
m^2_{s\bar{s}}(gH)  \; =  \; m^2_{s\bar{s}} \;  +  \;  \alpha \,  gH \; \; ,
\end{equation}
we get:
\begin{equation}
\label{4.15}
\Sigma_{s}(T=0,gH) \; \simeq  \;  \frac{1}{32 \, \pi^2}  \, \left \{ - \alpha \, \frac{gH}{m^2_{s\bar{s}} } \; - \;   
(1 \, + \alpha   \,   \frac{gH}{m^2_{s\bar{s}}} ) \ln  (1 \, + \alpha   \,   \frac{gH}{m^2_{s\bar{s}}} )     \right \} \; \; .
\end{equation}
As a matter of fact, we find that Eq.~(\ref{4.15}) allows to track quite well the data by using for $\alpha$ the value in
Eq.~(\ref{4.12}), but adopting a magnification factor of about 66.0.
\section{Chiral critical temperature}
\label{S5}
Previous studies for the SU(3) pure gauge theory~\cite{Cea:2005dg,Cea:2005td} and QCD with two degenerate staggered massive quarks with
 pion mass around 500 MeV~\cite{Cea:2007yv} in presence of an external constant background field showed that the critical deconfinement 
 temperature decreases with the chromomagnetic field strength according to:
\begin{equation}
\label{5.1}
T_c(gH) \; = \;  T_c(0) \; \left ( 1 \; - \; \frac{\sqrt{gH}}{\sqrt{gH_c}}  \right ) \; \; .
\end{equation}
From this last equation one infers that the deconfinement critical temperature decreases linearly with $\sqrt{gH}$ and eventually it vanishes at the critical chromomagnetic field strength $\sqrt{gH_c}$, leading to the remarkable color Meissner effect.
In both cases, $\sqrt{gH_c}$  turned out to be of order  1 GeV or, more precisely,     $\sqrt{gH_c} = 1.22(4)$ GeV in SU(3) 
and $\sqrt{gH_c} = 1.61(14)$ GeV in QCD$_2$.
Previous lattice studies, however, have been limited to relatively weak background fields, i.e. $gH < 1$ GeV$^2$. In addition,  from the results
in full QCD in external magnetic fields it is clearly evident that the dynamical quark masses strongly influences the behaviour
of the critical temperature on the external magnetic field strength.  This study addresses both limitations of previous work.  First, we performed simulations of (2+1)-flavor QCD at the physical point. Second, we considered background chromomagnetic fields up to $gH \simeq 7$ GeV$^2$. As discussed in the preceding Section,
the critical temperatures have been determined from the inflection point of the renormalised  chiral condensate.%
\begin{figure}
\includegraphics[width=0.95\textwidth,clip]{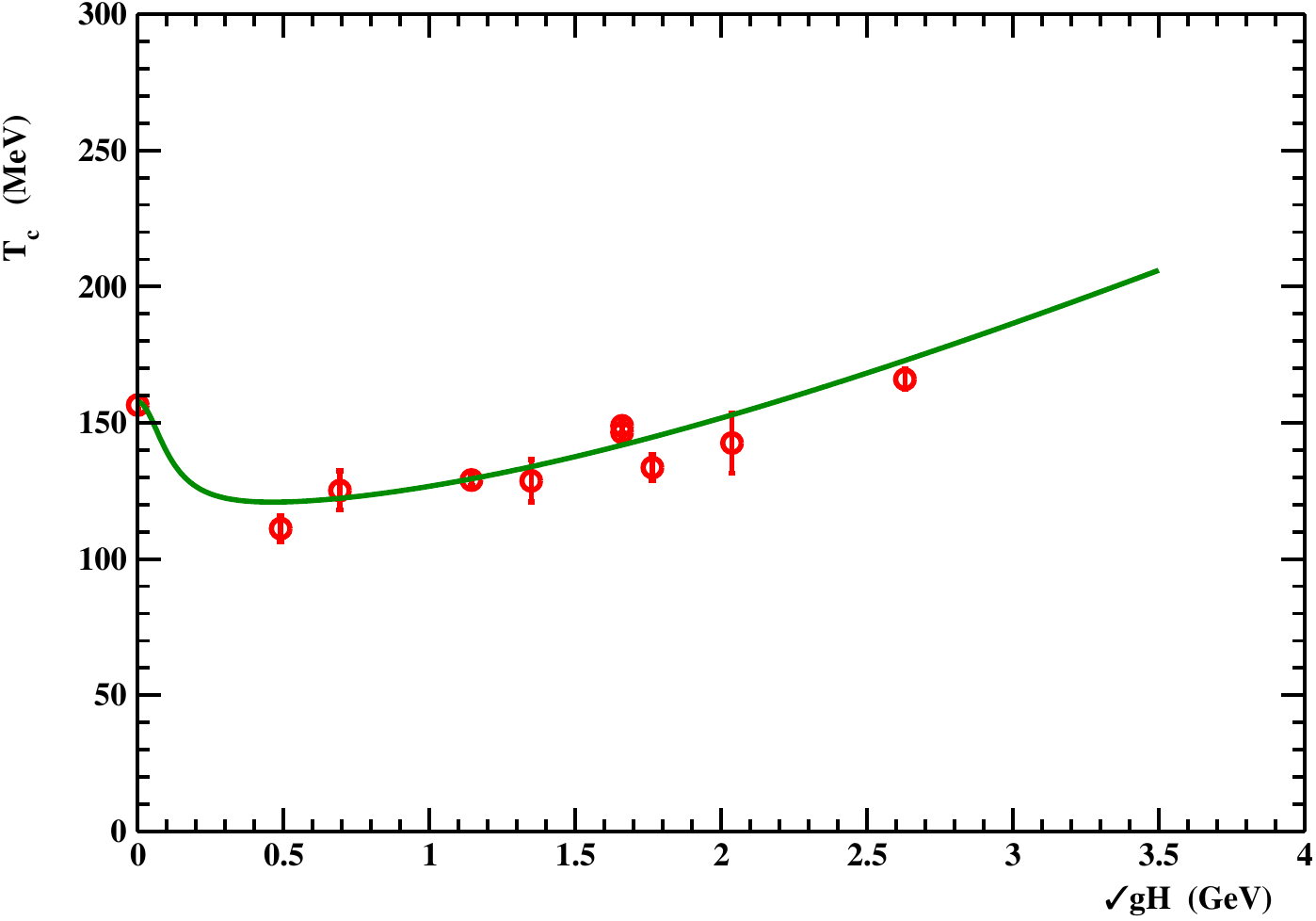}
\caption{\label{Fig7} 
The critical deconfinement temperature versus $\sqrt{gH}$. 
The continuous line is the best-fit of the lattice data  to Eq.~(\ref{5.2}). }
\end{figure}
In particular, the critical deconfinement temperature is assumed to be coincident with the pseudocritical temperature of the renormalised light 
chiral condensate. As a consequence, from Table~\ref{Table2} we infer the critical temperature $T_c$ as a function of $\sqrt{gH}$
as displayed in Fig~\ref{Fig7}.
Remarkably, Fig.~\ref{Fig7} shows a quite involved and unexpected behaviour of the critical temperature. At low chromomagnetic field
strengths, the critical temperature quickly decreases, in qualitative agreement with previous lattice studies. However, soon after it seems
that $T_c$ saturates to an almost constant value up to about $\sqrt{gH} \simeq 1.5$ GeV. After that, the critical temperature
begins to slowly increase almost linearly with the strength of the chromomagnetic field $\sqrt{gH}$.
The first interesting consequence is that full QCD at the physical point does not share the color Meissner effect, on the contrary
strong chromomagnetic fields reinforce the color confinement. \\
To gain insight into what is going on, we note that the dependence of the critical temperature on the strength of the chromomagnetic field
appears to be driven by two competing effects, the first tends to decrease $T_c$ while the second seems to reinforce
the confined phase. This leads us to try to reproduce $T_c(gH)$ with the following ansatz:
\begin{equation}
\label{5.2}
T_c(gH) \; = \;  T_c(0) \; \sqrt{ \frac{1 \; + \; a_1 \, gH \; + \; a_2 \, (gH)^2}{ 1 \; + \; a_3 \, gH}}  \; \; .
\end{equation}
The numerator within the square root should account for the increase of the critical temperature for strong enough chromomagnetic fields,
while the denominator takes care of the decrease of $T_c(gH)$ for small $\sqrt{gH}$. In addition we required that
asymptotically $T_c(gH) \, \sim \, \sqrt{gH}$.
Fitting the lattice data to Eq. (5.2) yielded the following results:
\begin{equation}
\label{5.3}
 T_c(0) \; =   \;  157.7 \; \pm \; 1.6 \; \text{MeV} \; \; ,
\end{equation}
\begin{equation}
\label{5.4}
a_1 \; = \; 52.0 \, \pm 3.1 \; \; , \; \;  a_2 \; = \; 9.0 \, \pm 1.3 \; \; , \; \; a_3 \; = \; 95.0 \, \pm 2.7 \; \; .
\end{equation}
Looking at Fig.~\ref{Fig7}  we see that our ansatz is able to track in a reasonable way ($\chi_r^2  \simeq 4$) the behaviour of
the critical temperature.  Our previous analysis of the zero-temperature light chiral condensate  (see Sect.4) suggested the presence of a
critical chromomagnetic field around 2 GeV where the pion mass goes to zero. However, the behaviour of the
critical temperature in Fig.~\ref{Fig7} does not manifest hints of critical chromomagnetic fields.  Further dedicated research is necessary to elucidate this matter fully.\\
Here, a comparison of our results with full QCD in external magnetic fields is essential.
\begin{figure}
\includegraphics[width=0.95\textwidth,clip]{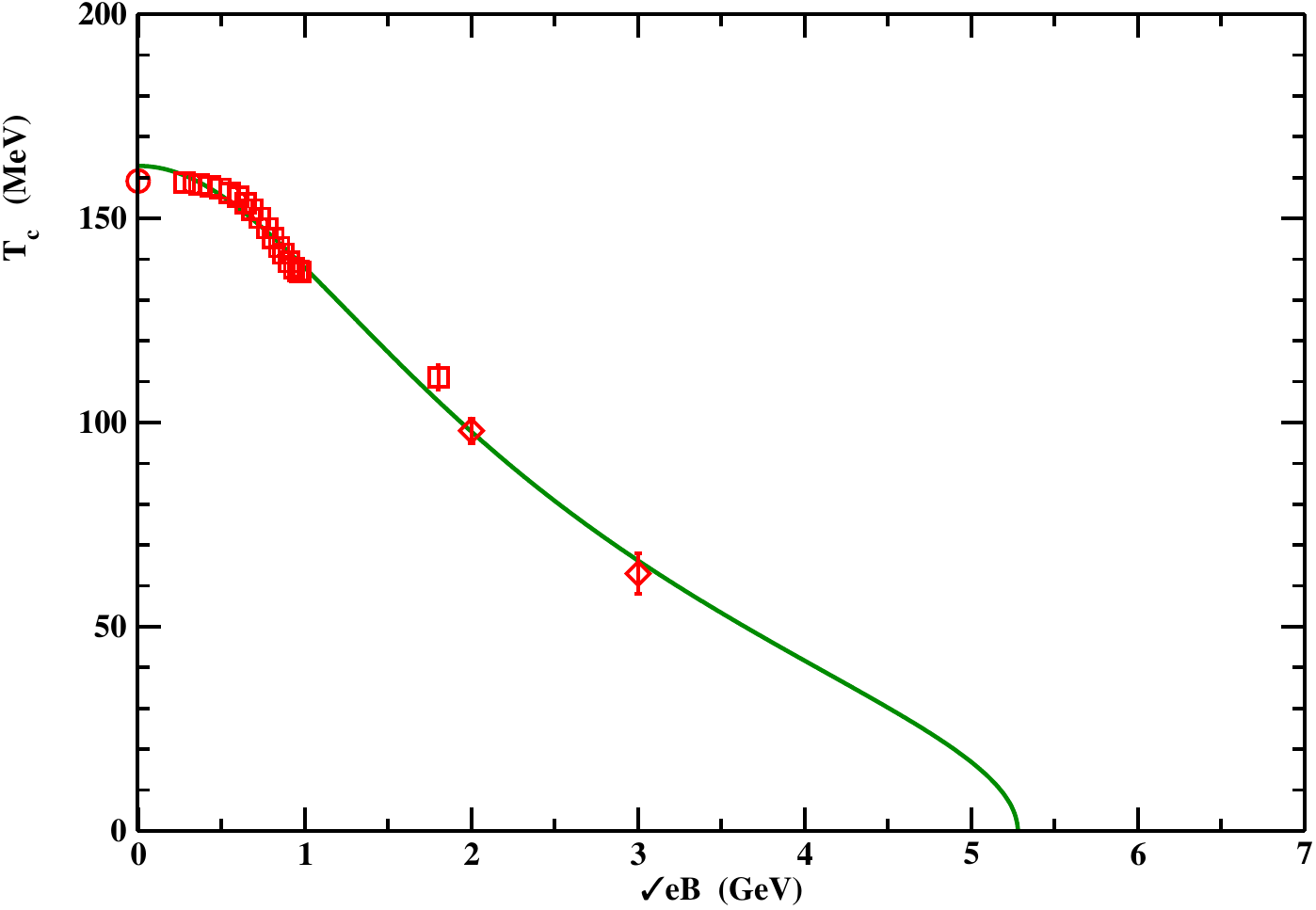}
\caption{\label{Fig8} 
The critical deconfinement temperature as a function of $\sqrt{eB}$.
The open circles and squares have been taken from Refs.~\cite{Bali:2011qj,Endrodi:2015oba};
the open diamonds are taken from Ref.~\cite{DElia:2021yvk}.
The continuous line is our best-fit of the lattice data  to Eq.~(\ref{5.2}).}
\end{figure}
Extensive non-perturbative lattice simulations using stout-improved staggered fermions have been conducted to investigate full QCD with (2+1) dynamical quark flavors at the physical point in the presence of external magnetic fields.
In Fig.~\ref{Fig8} we report the lattice data for the critical temperature versus $\sqrt{eB}$. The zero-field pseudocritical temperature:
\begin{equation}
\label{5.5}
 T_c(eB=0) \; =   \;  159.1 \; \pm \; 2.4 \; \text{MeV} \; 
\end{equation}
and the critical temperatures for $eB \, \lesssim 1$ GeV$^2$ have been taken from Ref.~\cite{Endrodi:2015oba}, while
the critical temperatures at $eB = 4,9$ GeV$^2$ are reported in Ref.~\cite{DElia:2021yvk}.
Looking at  Fig.~\ref{Fig8} we see that in the weak magnetic field regime the behavior of the critical temperature is quite similar
to the one in the case of external chromomagnetic fields. At very strong magnetic fields, the deconfinement temperature
exhibits a further decrease, which suggests the emergence of the ordinary Meissner effect, i.e. the vanishing of $T_c$ at a  critical magnetic field strength.
As a matter of fact, we tried to fit the lattice data to Eq.~(\ref{5.2}). The fitting procedure yielded very small values for the parameter $a_2$, which were always consistent with zero. To improve the accuracy of the remaining parameters, we fixed $a_2$ to zero and obtained the following results:
\begin{equation}
\label{5.6}
 T_c(0) \; =   \;  162.9 \; \pm \; 1.7 \; \text{MeV} \; \; ,
\end{equation}
\begin{equation}
\label{5.7}
a_1 \; = \; - \, 0.0359(235) \; \; , \; \;  a_2 \; = \; 0  \; \; ,  \; \; a_3 \; = \; 0.344(67) \;  \; \; .
\end{equation}
Remarkably, Figure~\ref{Fig8} shows that our Ansatz, Eq.~(\ref{5.2}), is able to track fair well the lattice data ($\chi^2_r \simeq 1$) across the entire range of applied magnetic field. Moreover, since $a_1$ is negative, it seems that there is a critical magnetic field
$\sqrt{eB_c} \simeq 5.3$ GeV leading to the Meissner effect. However, the whole effect is driven by the parameter $a_1$ that
is negative but very small and affected by a  large statistical uncertainty. Actually, from the statistical point of view we cannot esclude that
also the parameter  $a_1$ vanishes. In this case the deconfinement temperature goes to zero only asymptotically according to
$T_c \sim \frac{1}{\sqrt{eB}}$. \\
We saw that our Ansatz takes care of two competing effects of the external field on the deconfinement temperature. Comparing
Eq.~(\ref{5.7}) to Eq.~(\ref{5.4}) and considering that chromomagnetic and magnetic fields couple essentially in the same way to quarks,
we are led to the conclusion that the gluons are responsible for the increase of the critical temperature for strong enough
chromomagnetic fields. Likewise, also the decrease of $T_c$ for small external fields is mainly due to the vector gauge fields. This
explains naturally the smallness of the $a_1, a_2$ and $a_3$ parameters in the case of background magnetic fields since
magnetic fields are coupled to gluons via quark loops. In addition, thermal hysteresis that affected the chiral condensate
due to the presence of chromomagnetic fields must originate from the gluon dynamics. Indeed, the chiral condensate
in external magnetic fields did not displayed evidences of hysteresis. We shall return with further comments on this subject
later on. \\
As the last point addressed in the present Section we focus on the pseudocritical temperature $T_c^s$ extracted from the
inflection point of the renormalised strange chiral condensate.
\begin{figure}
\includegraphics[width=0.95\textwidth,clip]{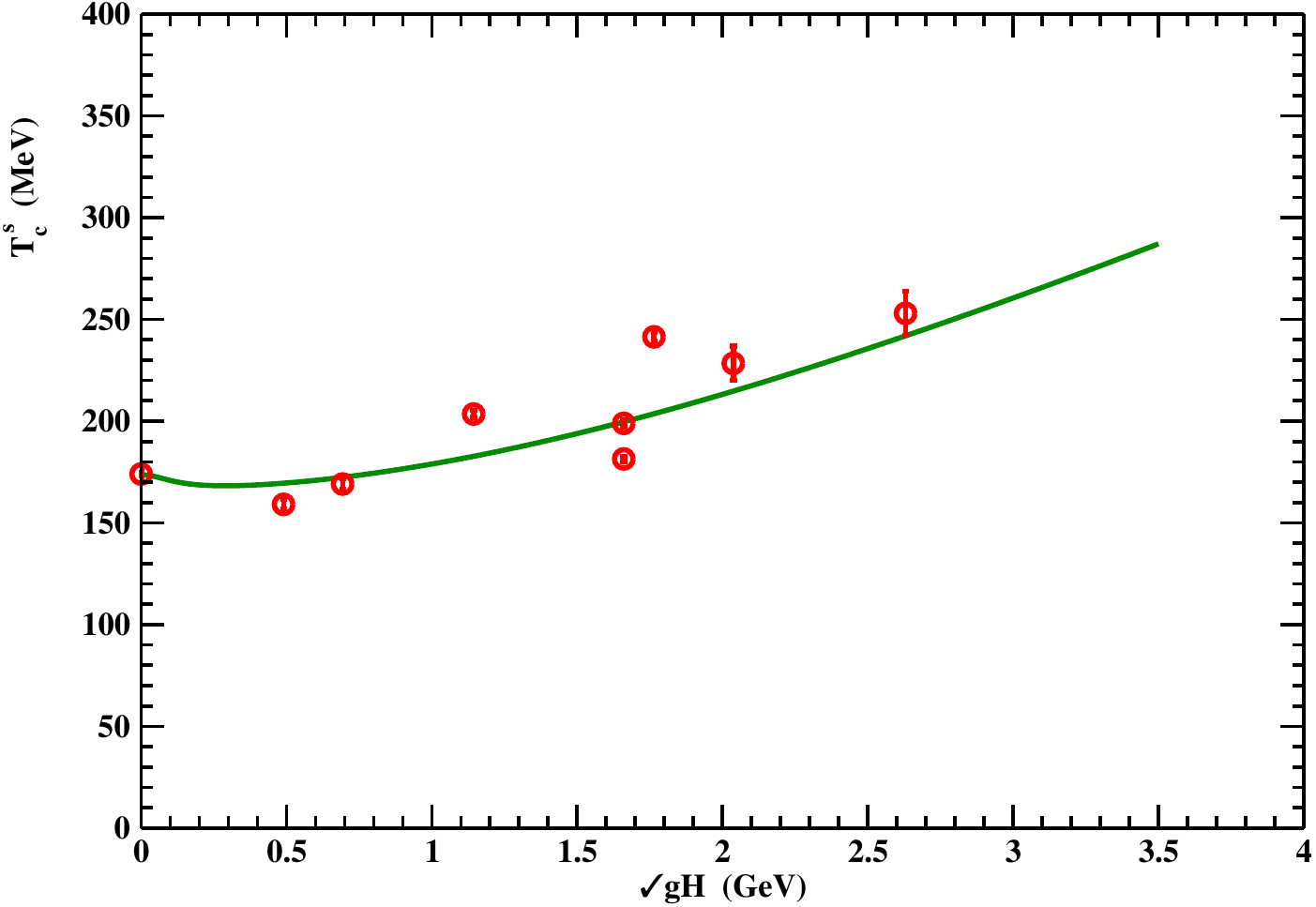}
\caption{\label{Fig9} 
The pseudocitical temperature $T_c^s$ versus $\sqrt{gH}$. The continuous line is the best-fit of the data  to Eq.~(\ref{5.2}). }
\end{figure}
In Fig.~\ref{Fig9} we report  $T_c^s$ (see Table~\ref{Table3}) as a function of the chromomagnetic field strength $\sqrt{gH}$. 
The zero-field pseudocritical temperature
\begin{equation}
\label{5.8}
 T_c^s  \; =   \;  174.0 \; \pm \; 2.0 \; \text{MeV} \; \; 
\end{equation}
has been determined in Ref.~\cite{Endrodi:2015oba} from the strange quark number susceptibility. Qualitatively,
Figure~\ref{Fig9} confirms that the strange pseudocritical temperature exhibits similar behavior to  the deconfinement temperature. It decreases in the weak field strength region, followed by a smooth increase towards strong field strengths.
The lattice data, however, are more scattered with respect to the previous cases. For this reason, to get a reliable fit to
Eq.~(\ref{5.2}) we constrained the fitting parameters $a_1$ and $a_2$ in the range allowed by Eq.~(\ref{5.4}).
In this way we obtain:
\begin{equation}
\label{5.9}
 T_c(0) \; =   \;  173.7 \; \pm \; 3.1 \; \text{MeV} \; \; ,
\end{equation}
\begin{equation}
\label{5.10}
a_1 \; = \; 55.1  \, \pm \, 5.2  \; \; , \; \;  a_2 \; = \;  9.0  \, \pm \, 1.0  \; \; ,  \; \; a_3 \; = \;  60.3  \, \pm \, 3.2    \;  \; \; .
\end{equation}
As expected, the resulting best-fit curve allows to track the lattice data only qualitatively.
\section{Summary and conclusions}
\label{S6}
The main aim of the present explorative work has been to investigate 
the properties of full QCD in external Abelian chromomagnetic fields by
means of non-perturbative numerical simulations on the lattice.
We have simulated QCD with (2+1)-flavour at the physical point using  HISQ fermion fields. The external chromomagnetic
field was implemented using the gauge-invariant Schr\"odinger functional at both zero and finite temperatures. We have investigated
the renormalized light and strange  chiral condensates across a broad temperature range around the pseudocritical temperature in the
presence of a uniform external Abelian chromomagnetic field with strengths varying  from 0.24 GeV$^2$ up to about 7.0 GeV$^2$.
In the critical region we have interpolated the lattice data with a functional form inspired to the Ising model. In this way, we was able
to estimate the pseudocritical chiral temperature and the zero-temperature renormalized chiral condensate.
Similar to external magnetic fields, we observe chromomagnetic catalysis within the confined phase of the gauge system
that turns out in the inverse catalysis in the high-temperature region. There are, however, important differences with respect to the results in lattice
studies of QCD in external magnetic fields. Indeed, in the confined phase, after taking into account that chromomagnetic fields cannot couple directly
to colorless hadrons, we suggested that the chromomagnetic catalysis arises from the fact that the external chromomagnetic field perturbs the
QCD quantum vacuum such as to change the spectra of the low-lying excitations. Accordingly, we showed that the behavior of the zero-temperature
renormalized chiral condensate versus the strength of the chromomagnetic field can be accounted for, albeit quite qualitatively, if the masses of the
lowest pseudoscalar mesons decrease linearly with $|gH|$. This leads us to suspect the presence of a critical chromomagnetic field with
$\sqrt{gH_0} \simeq 2.1$ GeV where the pions become massless.  It is not yet clear to us the meaning of this possible critical field, even
though it could signal a partial restoration of the chiral symmetry. In any case, we stressed that this last point needs to be addressed more carefully
by future dedicated lattice studies. \\
The most surprising result of our lattice simulations is  the ubiquitous presence of hysteresis affecting the renormalized light and strange
chiral condensate. The absence of hysteresis observed in lattice studies of full QCD at the physical point in presence of magnetic fields strongly suggests
that the gluons are playing the dominant dynamical role in the  instauration of hysteresis. This could account for the observed dependence of the critical deconfinement temperature on the chromomagnetic field strength that deviates significantly from the case of QCD in a magnetic field.
Indeed, the deconfinement temperature decreases in the weak field region, after that displays a broad plateau and finally it smoothly increases
for strong chromomagnetic field strengths. Our comparison with full QCD in external magnetic fields suggests that this peculiar behavior of the deconfinement temperature in the strong field regime can be attributed to gluon dynamics under strong chromomagnetic fields. \\
This work presents the first investigation of full QCD at the physical point in the presence of external Abelian chromomagnetic fields. The most surprising and unexpected result is the observation of hysteresis in both light and strange chiral condensate and the 
increase of the deconfinement temperature in the strong field region. Regarding hysteresis, it is known since long time that in magnetic materials
it is naturally connected to the presence of magnetic domains.
This suggests that a straightforward  interpretation of our results could imply the presence of chromomagnetic domains in the QCD vacuum. In this regard, 
it is worthwhile to mention that in a recent paper~\cite{Cea:2023} it was presented an attempt to derive from first principles the QCD quantum vacuum.
The resulting vacuum wavefunctional resembles a collection of chromomagnetic domains,  where each domain is characterized by a chromomagnetic condensate
pointing in arbitrary space and color directions. Color confinement arises from the lack of long-range correlations and the presence of a mass gap of 
the order of the inverse of the domain linear size.  The presence of chromomagnetic domains can explain the complex landscape
of the free energy leading to hysteresis. On the other hand, the paramagnetic behavior of gluons implies that in an external chromomagnetic
field it is expected an enhancement of the vacuum chromomagnetic condensate that, in turns, tends to increase the mass gap and, thereby,
the deconfinement temperature. At the same time, an external chromomagnetic field tends to polarize the domains resulting into a decrease of
the mass gap~\cite{Cea:2023}. These arguments motivated the Ansatz we introduced in Sect.~\ref{S5} to explain the behavior
of the deconfinement temperature as a function of chromomagnetic and magnetic field strengths. \\
In conclusion, aside from these last considerations, the results presented in this paper highlight important and unexpected
shared properties of the quantum vacuum of full QCD at the physical point.
\begin{acknowledgments}
We would like to thank Massimo D'Elia and Gergely Endrodi for their insightful comments and suggestions.
This investigation was in part based on the MILC collaboration's public lattice gauge theory code (\url{https://github.com/milc-qcd/}). Numerical calculations have been made possible through a Cineca-INFN agreement providing access to HPC resources at Cineca.
The authors acknowledge the HPC RIVR consortium (www.hpc-rivr.si) and EuroHPC JU (eurohpc-ju.europa.eu) for funding this research by providing computing resources of the HPC system Vega at the Institute of Information Science (www.izum.si), and EuroHPC (grant EHPC-BEN-2023B06-040). The authors acknowledge support from INFN/NPQCD project. 
This work is (partially) supported by ICSC – Centro Nazionale di Ricerca in High Performance Computing, Big Data and Quantum Computing, funded by European Union – NextGenerationEU.
\end{acknowledgments}


%

\end{document}